\documentclass[aps, prl,
twocolumn, %showpacs,
floatfix,
preprintnumbers,
nofootinbib,
superscriptaddress]
{revtex4_0}
\pdfoutput=1
\usepackage{natbib}
\usepackage{multirow}
\usepackage{graphicx, rotating,amssymb}
\ifx\pdfoutput\undefined
\usepackage[dvips,bookmarks]{hyperref}	% This is for arXiv.org
\else
\usepackage{hyperref}	% This is for pdftex
\fi
\hypersetup{colorlinks,bookmarksopen,bookmarksnumbered,citecolor=rossoc,
linkcolor=verdes,pdfstartview=FitH,urlcolor=rosso}

\usepackage{color}
\definecolor{rosso}{cmyk}{0,1,1,0.4}
\definecolor{rossos}{cmyk}{0,1,1,0.55}
\definecolor{rossoc}{cmyk}{0,1,1,0.2}
\definecolor{blu}{cmyk}{1,1,0,0.3}
\definecolor{blus}{cmyk}{1,1,0,0.6}
\definecolor{bluc}{cmyk}{1,1,0,0.1}
\definecolor{verde}{cmyk}{0.92,0,0.59,0.25}
\definecolor{verdec}{cmyk}{0.92,0,0.59,0.15}
\definecolor{verdes}{cmyk}{0.92,0,0.59,0.4}

\newcommand{\beq}{\begin{equation}}
\newcommand{\eeq}{\end{equation}}

\usepackage{lscape,xspace}

\def\AMS{{\sc Ams-02}\xspace}

\def\CALET{{\sc Calet}\xspace}

\def\DAMPE{{\sc Dampe}\xspace}
\def\ISSCREAM{{\sc Iss-Cream}\xspace}

\usepackage{twoopt}
\makeatletter

\newcommandtwoopt{\citeads}[3][][]{\href{http://adsabs.harvard.edu/abs/#3}
    {\def\hyper@linkstart##1##2{}
    \let\hyper@linkend\@empty\citealp[#1][#2]{#3}}}

\newcommandtwoopt{\citepads}[3][][]{\href{http://adsabs.harvard.edu/abs/#3}
    {\def\hyper@linkstart##1##2{}
    \let\hyper@linkend\@empty\citep[#1][#2]{#3}}}

\newcommandtwoopt{\citetads}[3][][]{\href{http://adsabs.harvard.edu/abs/#3}
    {\def\hyper@linkstart##1##2{}
    \let\hyper@linkend\@empty\citet[#1][#2]{#3}}}

\makeatother
\begin{document}
\title{Indications for a High-Rigidity Break in the Cosmic-Ray Diffusion Coefficient}

\author{Yoann G\'enolini}
\email{yoann.genolini@lapth.cnrs.fr}
\affiliation{Laboratoire d'Annecy-Le-Vieux de Physique Th\'eorique (LAPTh), Universit\'e Savoie Mont Blanc \& CNRS, 74941 Annecy Cedex, France}

\author{Pasquale D. Serpico}
\email{serpico@lapth.cnrs.fr}
\affiliation{Laboratoire d'Annecy-Le-Vieux de Physique Th\'eorique (LAPTh), Universit\'e Savoie Mont Blanc \& CNRS, 74941 Annecy Cedex, France}

\author{Mathieu Boudaud}
%\email{boudaud@lpthe.jussieu.fr}
\affiliation{Laboratoire de Physique Th\'eorique et Hautes \'Energies (LPTHE), UMR 7589 CNRS \& UPMC, 4 Place Jussieu, F-75252 Paris -- France}

\author{Sami Caroff}
%\email{caroff@llr.in2p3.fr}
\affiliation{Laboratoire Leprince-Ringuet (LLR), Ecole Polytechnique, CNRS/IN2P3, F-91128 Palaiseau, France}

\author{Vivian Poulin}
%\email{vivian.poulin@lapth.cnrs.fr}
\affiliation{Laboratoire d'Annecy-Le-Vieux de Physique Th\'eorique (LAPTh), Universit\'e Savoie Mont Blanc \& CNRS, 74941 Annecy Cedex, France}
\affiliation{Institute for Theoretical Particle Physics and Cosmology (TTK), RWTH Aachen University, D-52056 Aachen, Germany}

\author{Laurent Derome}
\affiliation{Laboratoire de Physique Subatomique et Cosmologie (LPSC), Universit\'e Grenoble Alpes, CNRS/IN2P3, 53 avenue des Martyrs, 38026 Grenoble, France}

\author{Julien Lavalle}
%\email{lavalle@in2p3.fr}
\affiliation{Laboratoire Univers \& Particules de Montpellier (LUPM),
  CNRS \& Universit\'e de Montpellier (UMR-5299),
  Place Eug\`ene Bataillon,
  F-34095 Montpellier Cedex 05 -- France}

\author{David Maurin}
\affiliation{Laboratoire de Physique Subatomique et Cosmologie (LPSC), Universit\'e Grenoble Alpes, CNRS/IN2P3, 53 avenue des Martyrs, 38026 Grenoble, France}

\author{Vincent Poireau}
%\email{poireau@lapp.in2p3.fr}
\affiliation{Laboratoire d'Annecy de Physique des Particules (LAPP), Universit\'e Savoie Mont Blanc \& CNRS, 74941 Annecy Cedex, France}

\author{Sylvie Rosier}
%\email{rosier@lapp.in2p3.fr}
\affiliation{Laboratoire d'Annecy de Physique des Particules (LAPP), Universit\'e Savoie Mont Blanc \& CNRS, 74941 Annecy Cedex, France}

\author{Pierre Salati}
%\email{salati@lapth.cnrs.fr}
\affiliation{{LAPTh}, Universit\'e Savoie Mont Blanc \& CNRS, 74941 Annecy Cedex, France}

\author{Manuela Vecchi}
%\email{manuela.vecchi@ifsc.usp.br}
\affiliation{Instituto de F\'isica de S\~ao Carlos, Universidade de S{\~a}o Paulo, CP 369, 13560-970,  S{\~a}o Carlos, SP, Brazil}

\preprint{
LAPTH-023/17,
LUPM:17-011
}
 \date{\today}
\begin{abstract}
Using cosmic-ray boron to carbon ratio (B/C) data recently released by the \AMS\ experiment, we find indications ({\it decisive evidence}, in Bayesian terms) in favor of a diffusive propagation origin for the broken power-law spectra found in protons ($p$) and helium nuclei (He). The result is robust with respect to currently estimated uncertainties in the cross sections, and  in the presence of a small component of primary boron, expected because of spallation at the acceleration site. Reduced errors at high energy as well as further cosmic ray nuclei data (as absolute spectra of C, N, O, Li, Be) may  definitively confirm this scenario.
 \end{abstract}

\maketitle

{\it \bf Introduction ---}
The  multiple deflections of cosmic rays (CRs)  on magnetic irregularities  cause their propagation to be a diffusive process. This increases their residence time in the Galaxy, and so their interaction probability with the interstellar medium (ISM). Their collision products include species which are otherwise rare or absent in the ISM, such as \textit{ ``fragile''} elements like Li-Be-B or antiparticles, like antiprotons ($\bar{p}$). These so-called secondary species {\bf({\sf SS})} have long been used to set constraints on propagation parameters in the generalized diffusion-loss equations linking the CR injection to the observable fluxes at Earth \cite{1950PhRv...80..943B}. Once tuned to measurements, these models define the framework within which other astroparticle physics investigations are performed, like indirect searches for dark matter via their
charged (anti)particle annihilation (or decay) byproducts.

The last decade has witnessed a major improvement in the precision and dynamical range of direct CR measurements, culminating with the AMS-02 experiment on board the ISS. Traditional theoretical models are under  strain, when challenged to match the precision of recent observations. On the one hand, the experimental error bars have shrunk to such a level that the stochastic nature of the sources  provides an irreducible limitation to theoretical predictive power
(see Ref.~\cite{Genolini:2016hte} for a recent study in that sense). On the other hand, the observations have revealed subtle features demanding an explanation, such as the broken power-law spectra  in $p$~\cite{Aguilar:2015ooa} and He fluxes~\cite{Aguilar:2015ctt}, and also probably present in heavier nuclei, confirming earlier indications by PAMELA~\cite{Adriani:2011cu} and CREAM~\cite{Ahn:2010gv}.
Theoretical studies should thus aim at reducing (or at least  assessing) uncertainties, while enlarging the range of phenomena to explain, i.e., should do ``more and better''.  For instance, a number of explanations have been put forward for the broken power laws. As reviewed in~\cite{Serpico:2015caa} (see also \cite{2012ApJ...752...68V}), the most promising tool to distinguish between different classes of models resides in the study of {\sf SS}, or alternatively of the ratio of a {\sf SS}, like B, to a (mostly) primary one, like C. If breaks are already present in the spectra accelerated at the sources, such a ratio should appear featureless, since the daughter nucleus ``inherits''  its parent features. If these features are due to propagation (as suggested by the similar rigidity at which it is seen in different species) the effect should be  twice as pronounced in {\sf SS}, thus emerging in a secondary over primary ratio, provided that a sufficiently high-precision measurement extending up to high rigidities is available.  The ratio of B/C fluxes recently released by \AMS\ \citepads{2016PhRvL.117w1102A} up to $\sim$2 TV provides such an opportunity.  Note that nontrivial features in B/C may be also due to the so-called \textit{distributed} reacceleration (DR) process (see~\cite{1987ApJ...316..676W} for an early proposal).
Although DR may have interesting implications for fine details of CR spectra \cite{2003A&A...410..189B, 2012A&A...544A..16T, 2017MNRAS.471.1662B}, the idea to use it to explain $p$ and He breaks \cite{2014A&A...567A..33T} is not viable due to qualitative and quantitative problems, as pointed out e.g. in~\cite{2015JPhG...42g5201E, 2017MNRAS.471.1662B}. {Furthermore, its impact on secondary fluxes seems to be negligible compared to spallation processes at source (see e.g.\citepads{2014ApJ...791L..22B}) which are discussed in the following. Yet, it has been recently argued in Ref.~\cite{2017MNRAS.471.1662B} that even if prominent DR is present, {\sf SS} spectra still show the need for an extra break, likely induced by diffusion. For these reasons, we exclude DR from the class of models tested hereafter.}

In this study, we investigate several hypotheses with the USINE code, in the limit  of a 1D diffusion model \cite{Jones2001}. This geometry is sufficient to capture the physics encoded in the B/C ratio;  the simplicity of the model is an asset to test the diffusion coefficient break and the robustness of our conclusions.
We follow a strategy which is complementary to recent trends: we restrict the theoretical framework to a sufficiently simple scenario with few fit parameters, and compare different hypotheses without introducing additional ones.
 For this purpose, we make use of break parameters determined by the $p$ and He analysis from \AMS, thus performing a test ``a priori''.

{\it \bf\label{background}Methodology ---}
Within a very large class of models, CR fluxes observed at Earth in the high-rigidity regime (tens of GV to hundreds of TV) are expected to depend mainly on the source term and the diffusion properties. Moving toward lower energies, additional effects  enter, such as convective winds, reacceleration, solar modulation, and energy losses. Given our primary goal to isolate features in the (effective homogeneous and isotropic) diffusion coefficient $K=K(R)$, we focus in the following on the rigidity range above ${\cal O}$(10) GV and keep as primary fit parameters its normalization $K_0$ and power-law index $\delta$. We also fix the diffusive halo height $L$ to 10 kpc, since it is a parameter largely degenerate with $K_0$.
We emphasize that our goal here is not to find ``the best fit'' parameters for the description of the data over the whole energy range, but identify and use the key physical variables on which the high-rigidity data depend. In this context, we test for two models, with the same number of free parameters. The conventional diffusion model
\begin{equation}
K(R)=K_0\,\beta\,(R/{\rm GV})^{\delta}\,,\label{K}
\end{equation}
vs
\beq
K(R)=K_0\,\beta\;\frac{(R/{\rm GV})^{\delta}}{\left\{1+\left(R/R_b\right)^{\Delta\delta/s}\right\}^{s}}\;\label{brokenK}
\eeq
where $s, \Delta \delta,$ and $R_b$ are, respectively, the smoothing, the magnitude and the characteristic rigidity of the break. These parameters are not extra parameters adjusted to the {\rm B/C} data, but result from a fit on the breaks in the \AMS $p$ and He spectra. In practice, we treat the break parameters  as {\em nuisance parameters}, whose best fit values and errors are extracted from $p$ and He. To do such an analysis correctly, it is necessary to take into account degeneracies between the parameters. This could be done thanks to their covariance matrix, which is unfortunately not provided by the \AMS collaboration. Hence, we perform a new {\em simultaneous} fit to the $p$ and He data, taking into account statistical and systematic uncertainties as described in Ref.~\cite{SamiPhD}.  Our results ($R_b=312^{+31}_{-26}\,$GV, $\Delta\delta=0.14\pm0.03$, $s=0.040\pm0.015$) are consistent with both sets of values found by \AMS,  and we checked that adopting the best fit values found in their publication would not affect our conclusions.
Note that the hypothesis~(\ref{brokenK}) means attributing the breaks in $p$ and He to diffusion. There are several proposals in the literature to produce such a behavior with microphysical mechanisms (e.g.~\cite{Blasi:2012yr}) or more complicated geometries and functional forms for $K(R)$~\cite{Tomassetti:2012ga}. The role played by the velocity parameter $V_a$ (as implemented in USINE \cite{2010A&A...516A..66P}) and the convective speed $V_c$ lessens as the rigidity increases. For instance, the data prove to be insensitive to the convective velocity $V_c$ as the fit yields a result consistent with zero when limited to higher and higher $R$. However, because of parameter degeneracies, we treat $V_a$ and $V_c$ as nuisance parameters whose variation range (from 0 to 10 km/s) is estimated via a preliminary fit over the full {\rm B/C} data.  We treat the solar modulation in the force field approximation, setting the Fisk potential to 0.730 GV, the average value over \AMS data taking period~\cite{GHELFI2016}---as retrieved from the online tool CRDB (\url{http://lpsc.in2p3.fr/crdb/}). We work in a 1D approximation since, apart from a renormalization in the effective value of the diffusion parameters (and particularly $K_0$), moving to a 2D geometry leads to similar fitting performances
~\cite{Genolini:2015cta,2010A&A...516A..66P}. We checked that, assuming a different low-$R$ dependence of the diffusion coefficient ($K\propto \beta^0$ instead of $\beta^1$ as discussed in \cite{2010A&A...516A..67M}), does not affect the statistical significance of the results obtained below. Needless to say, the best-fit values of the propagation parameters such as $\delta$ {\it depend} on the theory framework (and the range of $R$) one is fitting to \cite{2010A&A...516A..67M}, and we warn the readers that a comparison of the parameters obtained in our setup with similar parameters obtained in parametrically extended global fits to all data would be misleading, just like an extrapolation of our model to very low $R$, where it is expected a priori to fail.

The other ingredient upon which the results depend is the source spectrum. Boron is often considered as fully secondary, mostly produced by spallation of O and C. Fortunately, there is virtually no dependence of the B/C ratio upon the spectral shape of the primary C and O spectra~\cite{2002A&A...394.1039M, Genolini:2015cta}, at least as long as they are similar, which seems to be confirmed anyway from preliminary \AMS {}  data.
For definiteness, we fixed the injection power-law index to 2.1, but the specific choice is not essential. Hence, the main uncertain input determining the B source term, and thus the the transport parameters, are the spallation cross sections~\cite{2010A&A...516A..67M,Genolini:2015cta}. We compare our results for two choices, the GALPROP (GAL) data set \cite{2003ICRC....4.1969M} and Webber 2003 (W03) one \cite{2003ApJS..144..153W}. Since both  cross-section formulations assume a constant extrapolation above some energy, this comparison might not capture the whole uncertainty, in particular on the {\it shape} of the B/C in the energy range of interest. To assess its importance, we also test a different extrapolation, assuming a very mild growth of all cross sections with  $\ln^2 E$, $E$ being the energy per nucleon. This is certainly the reasonable leading growth behavior for the total and inelastic nucleon-nucleon cross section~\cite{2011PhRvL.107u2002B}, and leads to a corresponding growth in nucleus-nucleus collisions, as expected based on  Glauber models and experimentally checked in proton-air cross-section measurements in extensive air-shower detectors, see, e.g., Ref.~\cite{Gaisser:1986haa}. Lacking a more certain alternative, we further assume that the branching ratio into B is  $E$-independent. In practice we adopt for C, O, and B cross sections the same rise in {\bf $E$} as $\sigma_{pp}$  (see e.g \cite{2015PhRvD..92k4021B}), starting at the energy at which the total $pp$ cross section starts growing (zero derivative), i.e. around 100\,GeV/nuc (Lab frame). Continuity with the low-energy cross section is imposed. The resulting behaviour resembles---at least at visual inspection---the trends reported in the recent Monte Carlo study~\cite{Mazziotta:2015uba}.

Finally, the hypothesis that all B is secondary implicitly assumes that the acceleration time at the source is  small if compared to diffusion time $t_K\propto K^{-1}$. For a  time scale $t_A$ of a source capable of accelerating particles to $E\gtrsim\,$TeV/nuc,  one expects a primary to secondary fraction of B proportional to $t_A/t_K$.
For a gas density value typical of the ISM $n\sim 1$ cm$^{-3}$,  a C  nucleus interacts producing a B daughter with a probability of the order of $r\,\sigma\, n\, c\, t_A\sim 0.6\%$ for a cross section $\sigma\simeq 60\,$mb, where $r=4$ accounts for the standard strong shock compression factor and an active lifetime of $3\times 10^4$ yr is assumed. This is of the order of the age of the oldest supernova remnants detected in TeV $\gamma$-rays---hence capable of accelerating charged parent CRs to higher energy---such as the one in the W51 complex~\cite{2012A&A...541A..13A}.
 Accounting for the contribution to B by other nuclei, a benchmark value for primary B at the level of 1\% of C is reasonable and consistent with past publications, see, e.g., Eq.~(10) in~\cite{Aloisio:2015rsa}.

 It would also be  important to account for correlations between different energy bins,  usually captured by the correlation matrix. Lacking this information, we focus on two extreme cases: i) completely correlated systematics; ii) completely uncorrelated systematics. As this study is insensitive to global normalization factor{\bf s}, a good approximation for the former case is to use the statistical errors only ($\sigma_{\rm stat}$). For the latter case, the total uncertainty for each data point is defined as the quadratic sum of statistical and systematic uncertainties ($\sigma_{\rm tot}$). Note that a toy-correlation matrix can be constructed based on the detailed systematic errors in \cite{2016PhRvL.117w1102A} and a model of the energy correlations for each of these systematics. We checked that our qualitative results do not change using this toy model, although they indicate the quantitative importance of the covariance matrix, whose publication by {\bf \AMS} could prove very useful.

\begin{figure}[!th]
\centering
\includegraphics[width=0.8\columnwidth]{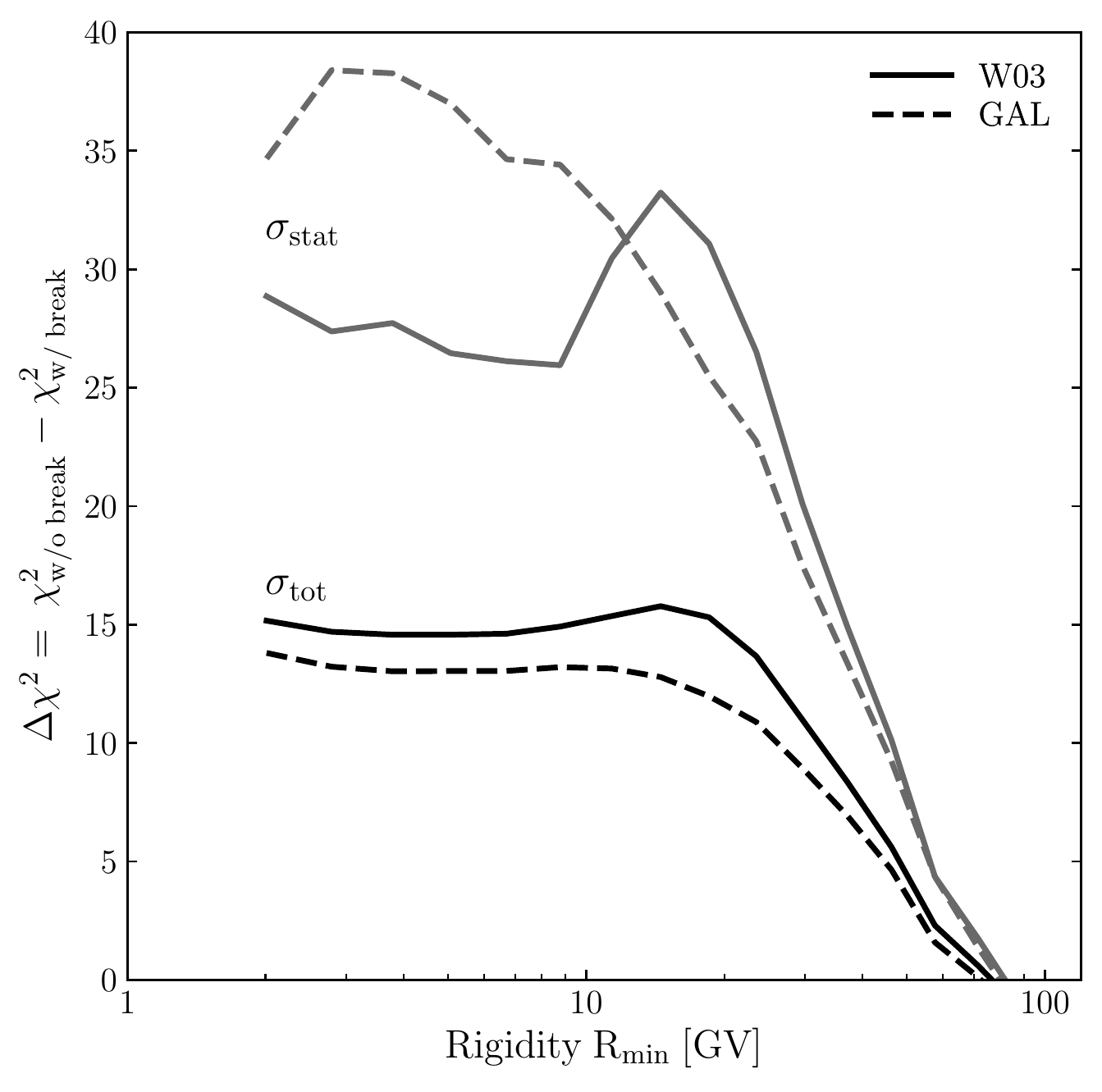}
\caption{\footnotesize Evolution of $\Delta\chi^2$ (with and without the break) vs the minimal rigidity $R_{\rm min}$ above which the fit is performed. Several cases are reported, using the GALPROP (GAL) or Webber 2003 (W03) cross-section data sets, and considering either statistical ($\sigma_{\rm stat}$) or total ($\sigma_{\rm tot}$) uncertainties.}\label{fig:delta_chi2}
\end{figure}

\begin{figure}[!th]
\centering
\includegraphics[width=0.9\columnwidth]{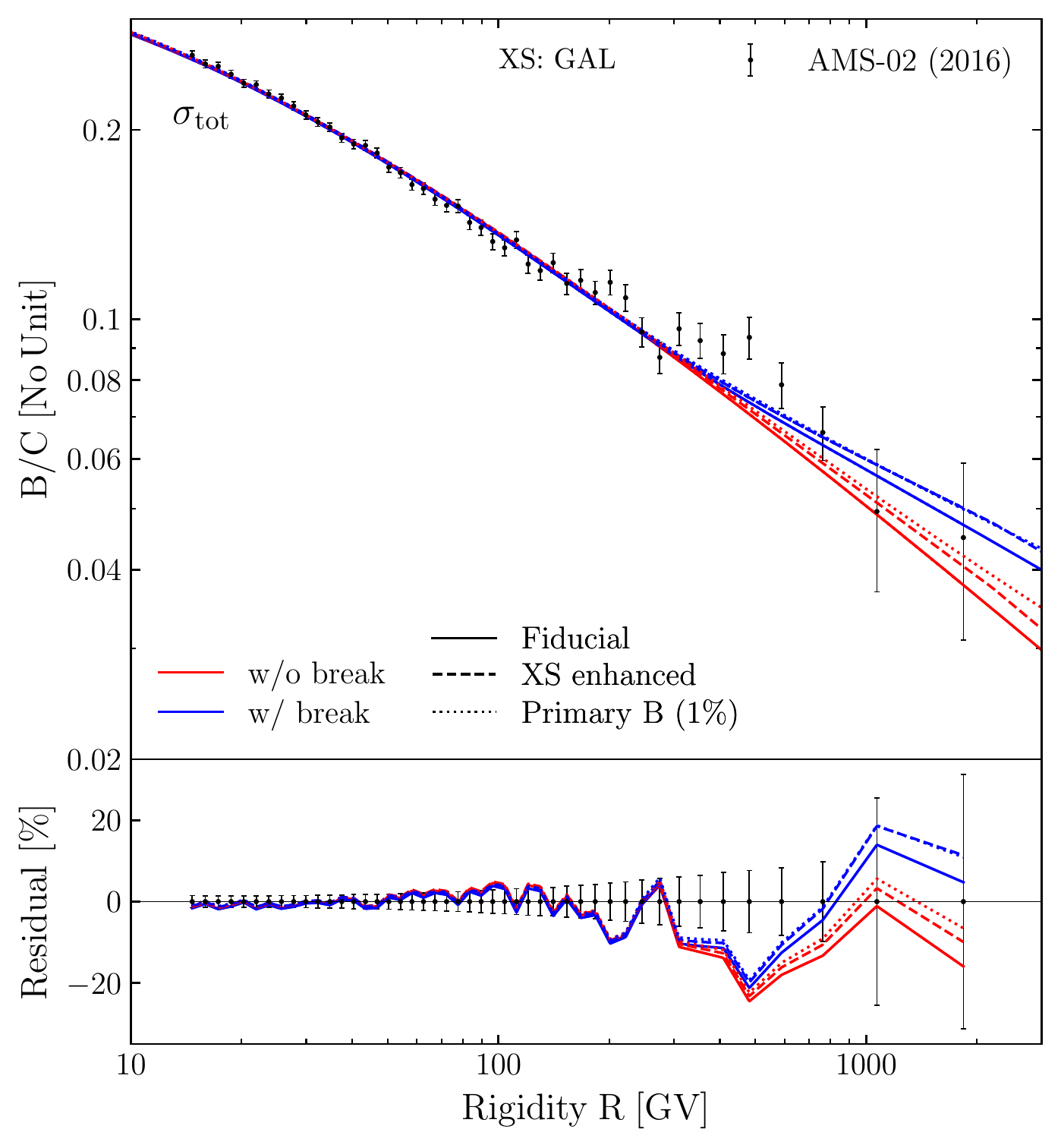}
\caption{\footnotesize Best fits and residuals with (blue) and without (red) the break using GALPROP cross sections and $\sigma_{\rm tot}$, for the different models considered in the text.}\label{fig:galprop_break}
\end{figure}

{\it \bf  Results---}
Since we focus on high-$R$, we fit the B/C data above $R_{\rm min}$, and gauge how the fit changes with  a break in the diffusion coefficient, calculating the $\Delta\chi^2$ between the best-fit obtained using  Eq.s~(\ref{brokenK}) and (\ref{K}). To check that the exact choice of $R_{\rm min}$ is not crucial we perform a scan on $R_{\rm min}$.   The $\Delta \chi^2$ vs $R_{\rm min}$ are plotted in Fig.~\ref{fig:delta_chi2}, for the Webber (solid line) and GALPROP (dashed line) cross-section formulations. In all cases, the fit improves when the break is introduced. As expected, a larger $\Delta\chi^2$ is found when  $\sigma_{\rm stat}$'s only are considered, although the nominal quality of the fit (in a frequentist approach) degrades. Within cross-section errors,  $\Delta\chi^2$ is approximately constant up to $R_{\rm min}=20$\,GV, and decreases above. This confirms that any  choice  $2\,{\rm GV}\lesssim R_{\rm min}\lesssim20$\,GV would lead to similar results of our test, while cutting at too high rigidities would hamper its statistical power since the baseline in $R$ becomes too short to highlight significant changes in the effective $\delta$.
Hypothesis tests are better performed by computing the {\em Bayesian evidence} $\kappa$ of the two models \cite{jeffreys1961theory,Kass95bayesfactors,2014bmc..book.....H}. In our  case
\begin{equation}
2\log(\kappa)=\Delta \chi^2\,,
\end{equation}
since both models share the same parameters,  and the ratio of the priors cancels in $\kappa$. In the conventional Jeffreys scale, a value of $2\log(\kappa) > 10$ is considered  ``decisive evidence'' \cite{jeffreys1961theory,Kass95bayesfactors,2014bmc..book.....H}. As shown below, in our analysis this criterion is always satisfied, for all assumptions tested (e.g. different choices for the spallation cross sections).
Of course, one may worry that other physical effects could imitate the break in the diffusion coefficient. We test the robustness of our model against two of them:  a) we include a different, but physically motivated high-energy extrapolation of the cross sections;  b) by adding a reasonable amount of primary B, corresponding to 1$\%$ of the C source term.
 Again, note that we do not extend our theory space with extra parameters to be fitted.
The best-fit values for each model are summarized in Tab.~\ref{tab:results} for $R_{\rm min}=15\,$GV.  We have checked in each case the independence of the results from the
exact choice of $R_{\rm min}$. In all cases the fits with break are better,  yielding a smaller $\chi^2$. The inferred $\delta$ is only altered by $\sim$0.01, well below the magnitude of the break. None of the potentially degenerate effects mentioned above significantly alters the $\Delta\chi^2$: the indication for the break remains ``decisive'' ($\Delta\chi^2\geq 10$). Figure~\ref{fig:galprop_break} displays the best fits reported in Table~\ref{tab:results}, using GALPROP spallation cross sections and $\sigma_{\rm tot}$. The residuals show the  weight of the six high-energy data points lying between 300\,GV and 800\,GV, stressing the importance of reducing the error bars there to tighten the test.

%%%
%DISCUSSION
%%%

{\bf Discussion and conclusions ---} By analyzing \AMS B/C data, we have found a ``decisive evidence'' (in a Bayesian sense) in favor of a high-rigidity break in the cosmic-ray diffusion coefficient, matching the similar features found in $p$ and He spectra. This suggests that the three observables ($p$, He, B/C) may find a {\it simultaneous} explanation for their spectral features in a model where the break is due to diffusion. We have conducted our study in a rather minimal theoretical setup, but we have tested the robustness of our conclusions with respect to effects such as the high-energy behavior of the cross sections or the presence of a reasonably small primary B component.

It is unclear at the moment if---in a frequentist approach---our results suggest that the underlying models are inadequate to describe the data. Overall, at least for GALPROP cross sections and for the analysis with $\sigma_{\rm tot}$, our fits with the break are of acceptable quality. The fit quality assuming  $\sigma_{\rm stat}$ is instead  poor. Lacking \AMS information on the error correlations, we may speculate that the actual situation is in between. Even then, it might still be that the simple models considered here provide an acceptable description at high-$R$: for instance, theoretical predictions are not error-free, but should be at the very least subject (via the primary C) to the kind of space-time source stochasticity effects first assessed in~\cite{Genolini:2016hte},  comparable to \AMS statistical uncertainties.

None of the conventional parameters in more extended theoretical models (like $V_c$, $V_a$, etc.) appears degenerate with the kind of high-$R$ feature discussed here. While their introduction is certainly important in attempts to explain the data over the whole range of $R$, it appears unlikely that those effects might significantly alter our conclusions, as confirmed by some preliminary tests.  One may be tempted to achieve a better fit by extending the model space with ``nonconventional'' free parameters, such as leaving either the diffusion break parameters or the primary B fraction free, as we have checked a posteriori. The consequences of a nominally better fit, however, are serious: allowing for a break significantly larger than the one found in $p$ and He (or a primary B fraction as high as 4.5\% of the C) would spoil the emerging global understanding of the broken power-law phenomenon. It may also raise additional problems, such as a significant overshooting of high-energy antiproton data (see the Appendix for an illustration of this tension).
We believe that a global understanding of the key features presented by CR data is preliminary to a detailed ``element-by-element'' modeling, if that is at all possible within current theoretical capabilities. In this spirit, a test of the ideas discussed here will probably benefit more of a first coherent understanding of an enlarged data set, including absolute flux measurements of primary species like C and O, ``intermediate'' ones like N, or secondary ones like Li, Be, B notably in the high-$R$ regime, rather than of a complete description of the B/C down to very small rigidities. Needless to say, future results from \AMS---including information on uncertainty correlations and/or high precision data covering even higher energies (e.g., \CALET~\cite{2017APh....91....1A} on ISS, the \DAMPE satellite~\cite{2017arXiv170105046G}, and \ISSCREAM~\cite{2014AdSpR..53.1451S} to be launched soon)---will be determinant.
\begin{table*}
\begin{tabular}{l l || c c c | c c c | c || c c c | c c c | c || c c c | c c c | c ||}
\hline \hline
   \multicolumn{2}{c}{Fit cases}  & \multicolumn{7}{c}{Fiducial}                                                                 & \multicolumn{7}{c}{Cross section enhanced}                           &    \multicolumn{7}{c}{Primary boron $Q_B/Q_C=1\%$}                                           \\
    Error         &     \multicolumn{1}{c}{Spal. XS}        & \multicolumn{3}{c}{w/o break}                    & \multicolumn{3}{c}{w/ break} & \multicolumn{1}{c}{~}  & \multicolumn{3}{c}{w/o break}      & \multicolumn{3}{c}{w/ break}  & \multicolumn{1}{c}{~}     & \multicolumn{3}{c}{w/o break}      & \multicolumn{3}{c}{w/break}     & \multicolumn{1}{c}{~}   \\
\hline
    \parbox[t]{2mm}{\multirow{4}{*}{\rotatebox[origin=c]{90}{$\sigma_{\rm stat}$}}}              &              & $K_0$  & $\delta$ & $\chi^2$ &            $K_0$  & $\delta$ & $\chi^2$        & $\Delta\chi^2$        &      $K_0$  & $\delta$ & $\chi^2$     &      $K_0$  & $\delta$ & $\chi^2$      &  $\Delta\chi^2$       &     $K_0$  & $\delta$ & $\chi^2$     &   $K_0$  & $\delta$ & $\chi^2$ & $\Delta\chi^2$        \\
                  &  W03               & $2.7$ & $0.67$ & $197$ &     $2.7$ & $0.68$ & $164$ & {\bf 33 }&         $2.7$ & $0.67$ & $190$ &            $2.7$ & $0.68$ & $160$ & {\bf 30 }&                $2.8$ & $0.69$ & $155$                    & $2.8$ & $0.69$ & $131$ & {\bf 24 } \\
                  &  GAL               & $4.3$ & $0.62$ & $160$ &     $4.3$ & $0.62$ & $131$ & {\bf 29 }&        $4.3$ & $0.62$ & $154$ &            $4.2$ & $0.62$ & $127$ & {\bf 27 }&               $4.4$ & $0.64$ & $126$                    & $4.3$ & $0.64$ & $105$ & {\bf 21 } \\
 \hline
\parbox[t]{2mm}{\multirow{2}{*}{\rotatebox[origin=c]{90}{$\sigma_{\rm tot}$}}}                  &  W03               & $4.5$ & $0.58$ & $84$  &     $4.3$ & $0.59$ & $68$ & {\bf 16 }&         $4.4$ & $0.58$ & $80$ &             $4.3$ & $0.59$ & $65$  & {\bf 15 }&                $4.4$ & $0.60$ & $69$                    & $4.2$ & $0.61$ & $57$ &{\bf 12 } \\
         &  GAL               & $7.4$ & $0.52$ & $62$  &     $7.1$  & $0.53$ & $50$ &{\bf 12 }&         $7.3$ & $0.52$ & $59$ &             $7.0$ & $0.53$ & $48$ & {\bf 11 }&                 $7.2$ & $0.54$ & $52$                    & $6.9$ & $0.55$ & $42$ & {\bf 10 }   \\
\hline
\end{tabular}
\caption{\footnotesize Best fit values for $K_0$ (in units of 10$^{-2}$~kpc$^2\,$Myr$^{-1}$) and $\delta$, using \AMS B/C data above $R_{\rm min}=15\;$GV. The number of degrees of freedom is $46-2=44$. For each case described in the Letter, we compare the best $\chi^2$ with and without the break. Two different spallation cross-section (Spal. XS) data sets are tested, i.e., GALPROP (GAL) and Webber (W03), as well as different choices for the data uncertainties. For guidance, typical best-fit errors in the $\sigma_{\rm stat}$ cases are of $1\%$ on $\delta$ and $2\%$ on $K_0$, whereas in the case of $\sigma_{\rm tot}$ they are of $2\%$ and $6\%$, respectively.}\label{tab:results}
\end{table*}

{\it \bf Acknowledgments ---} This work has been supported by the ``Investissements d’avenir, Labex ENIGMASS'', by the French ANR, Project No. DMAstro-LHC, ANR-12-BS05-0006, and by the CNES, France. We also acknowledge funding from University Savoie Mont Blanc under the AAP g\'enerique 2017 program. M.B. acknowledges support from the European Research Council (ERC) under the EU Seventh Framework Program (FP7/2007-2013)/ERC Starting Grant (agreement n. 278234 — NewDark project, PI: M. Cirelli). S.C. is supported by the ``Investissements d'avenir, Labex P2IO'' (ANR-10-LABX-0038) and by the French ANR (ANR-11-IDEX-0003-01). J.L. is supported by the OCEVU Labex (ANR-11-LABX-0060), the CNRS-IN2P3 Theory project {\em Galactic Dark Matter}, and European Union's Horizon 2020 research \& innovation program under the Marie Sk\l{}odowska-Curie grant agreements No 690575 and No 674896. M.V. is  grateful  to  the  S\~ao  Paulo  Research  Foundation  (FAPESP) for  the  support   through  grants  2014/19149-7 and 2014/50747-8.

%%%%%%
~\\ {\bf Appendix ---}Below we compute the {\it antiproton} signal associated to:
\begin{itemize}
\item[i)] one propagation model including the break in the diffusion coefficient and only a relatively small fraction of primary Boron (1\% of the primary Carbon flux), i.e. the case of the last column of Table I of the letter.
\item[ii)] a propagation model where the high-energy hardening in the B/C ratio is fully attributed to a primary B component, {\it fitted} to the data (hence, with an additional free parameter with respect to the models considered in the letter). This procedure results into a primary Boron fraction amounting to 4.5\% of the primary Carbon flux.
\end{itemize}
\begin{figure*}
\centering
\includegraphics[width=0.49 \linewidth]{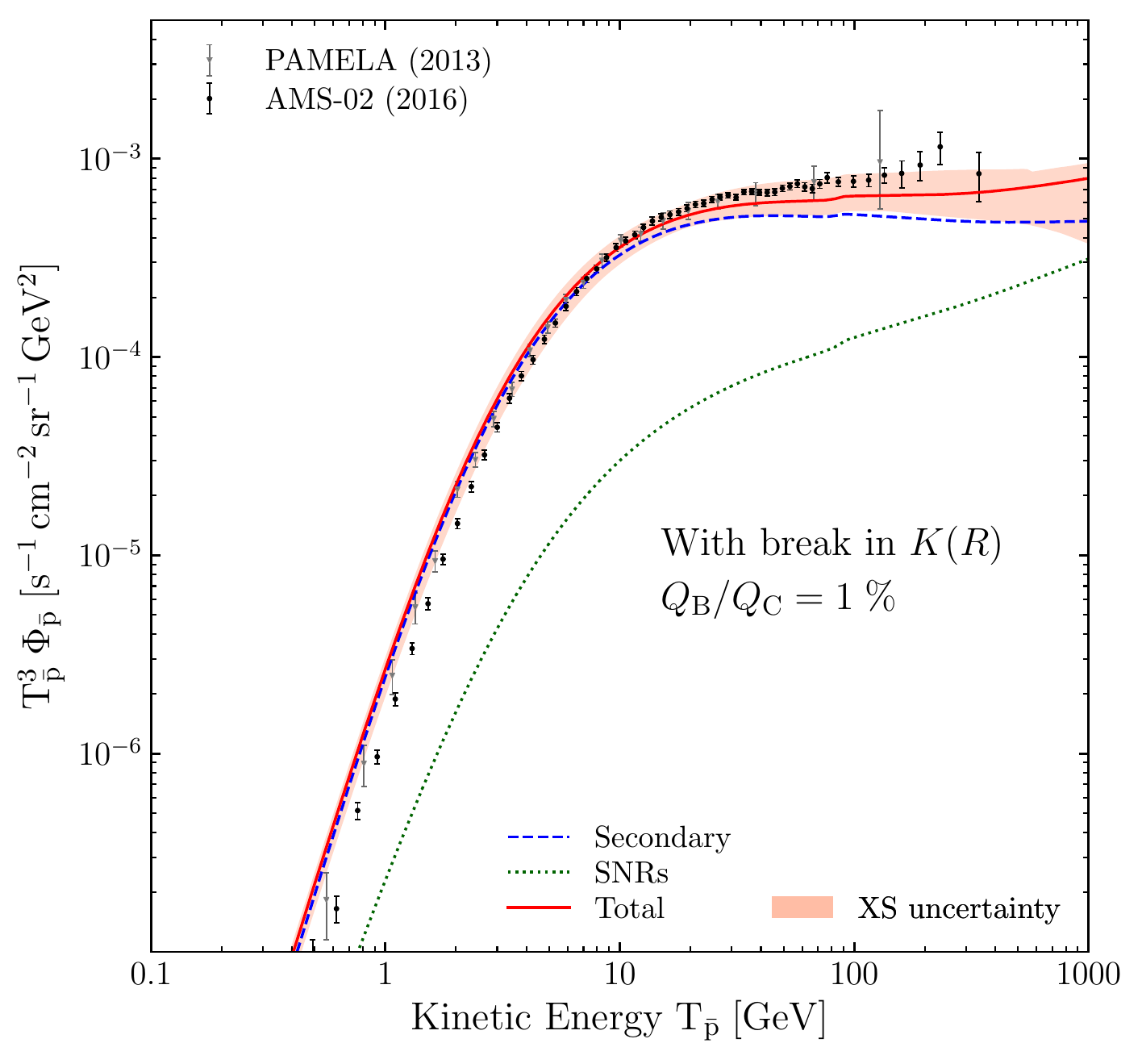} \includegraphics[width=0.49 \linewidth]{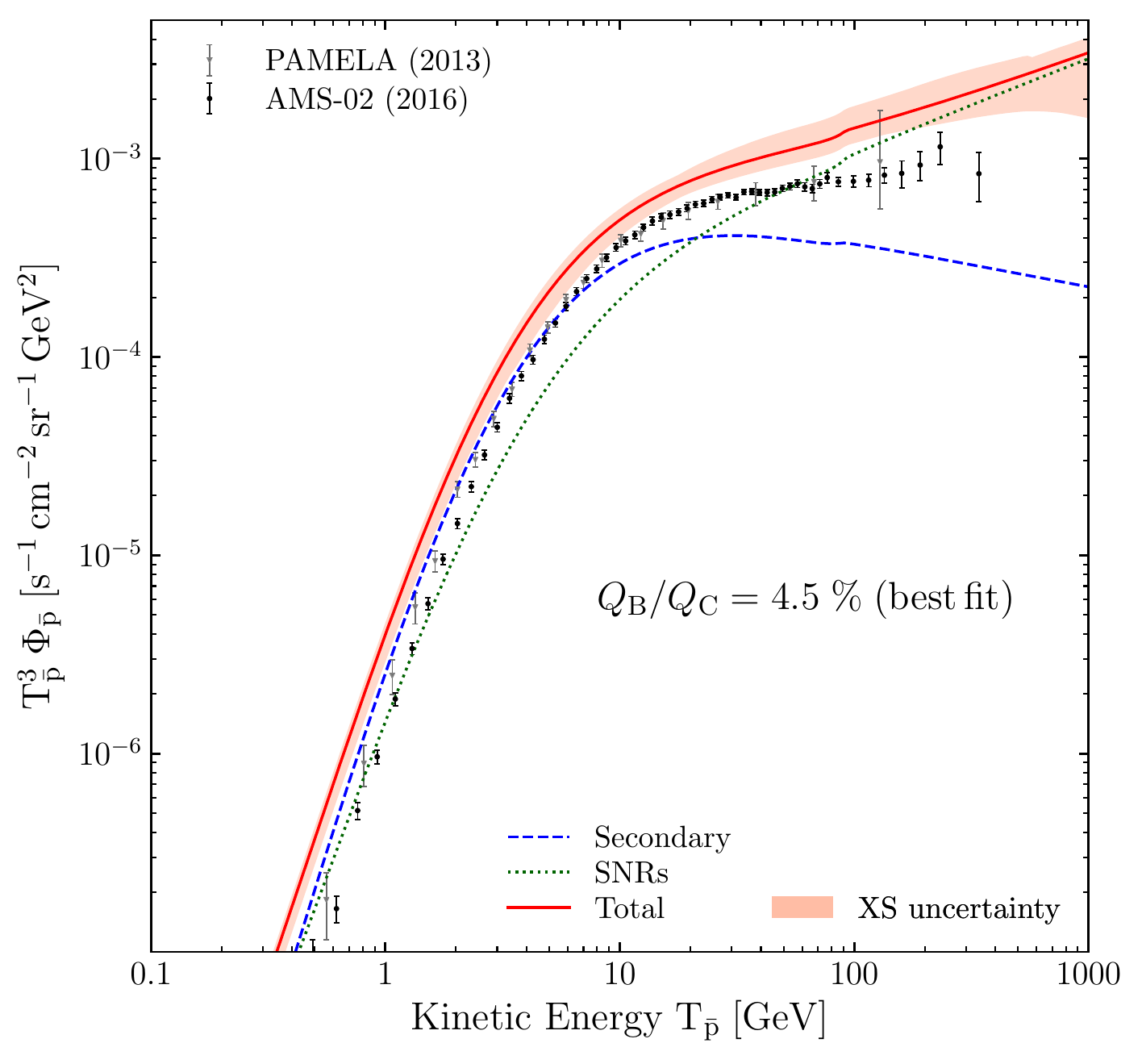}
\caption{Secondary antiprotons (dashed blue), primary antiprotons (dotted green) and total antiproton flux (solid red). \textbf{Left:} propagation parameters ($K_0 = 0.069$ kpc$^2$Myr$^{-1}$ and $\delta = 0.55$) obtained by fitting the B/C ratio including the break in the diffusion coefficient and a small fraction of primary Boron (1\% of the primary Carbon flux). \textbf{Right:} no break but a large fraction of primary Boron (4.5\% of the primary Carbon flux) fitted to the B/C data ($K_0 = 0.066$ kpc$^2$Myr$^{-1}$ and $\delta = 0.59$).}
\label{Fig:suppl1}.
\end{figure*}
For the B/C fits to AMS-02 data we are using the GALPROP cross-section dataset and add the statistical and systematic errors in quadrature (corresponding to GAL / $\sigma_{\rm tot}$ case in Table I). In both cases, the antiproton flux has a secondary contribution and a primary one, with the grammage at the source obviously corresponding to the assumed primary fractions in B/C.  We adopt a calculation procedure analogous to what is described in~\citepads{2015JCAP...09..023G}, with antiproton cross-section parameterization following~\citepads{2014PhRvD..90h5017D} and assuming a ratio $\bar{n}/\bar{p} = 1.3$. Uncertainties in these cross-sections are estimated to be at the level of 20-30\% in the range covered by data~\citepads{2014PhRvD..90h5017D}. We compare our predictions to the antiproton data borrowed from \citepads{2013JETPL..96..621A} and \citepads{2016PhRvL.117i1103A}.

From a simple visual inspection of Fig.~\ref{Fig:suppl1}, it is obvious that, while case i) provides a satisfactory agreement with the data within errors, case ii) seems excluded~\footnote{The discrepancies at low rigidity, present in both panels, should not be over-interpreted: the propagation model has only been fitted to B/C channel above the rigidity of 15 GV, and we are adopting a very simplistic treatment of the solar modulation, in terms of Fisk potential $\phi_F=560\,$MV, without reporting any associated uncertainty. }. Its tension with data appears hard to reconcile within uncertainties, since it involves both a sizable normalization mismatch and a different spectral slope than what inferred from measurements.

Although this preliminary calculation is not meant to replace an exhaustive study of the antiproton diagnostic potential, it is a clear example substantiating the claim reported in the main text that alternative solutions trying to avoid the introduction of a diffusive break are probably in tension with the data. In that respect, it should be noted that a primary boron contribution $Q_B/Q_C$ larger than 1\% (3\%) in the $\sigma_{\rm tot}$ ($\sigma_{\rm stat}$) case, may downgrade the \textit{decisive evidence} to a \textit{strong} one. Needless to say, such an exploratory cross-check may be extended to inspection of further cosmic ray species (such as Lithium) and observables (such as the anisotropy)  as well as to constrain additional physical effects like the amount of distributed re-acceleration in strong shocks, recently studied in Ref.\citepads{2017MNRAS.471.1662B}.
\bibliographystyle{ieeetr}
\bibliography{BC.vPRL}

\end{document}